\documentclass[letterpaper, 10 pt, conference]{ieeeconf}  
\IEEEoverridecommandlockouts                              

\usepackage{graphics} 
\usepackage{epsfig} 
\usepackage{amsmath} 
\title{\LARGE \bf
Multimodal Modeling of Ultradian Rhythms Using the Hankel Alternative View of Koopman (HAVOK) Analysis
}

\author{Emmanuel Molefi$^{1}$, Billy C. Smith$^{1}$, Christopher Thornton$^{1,2}$, Peter N. Taylor$^{1,3,4}$, Yujiang Wang$^{1,3,4}$% <-this % stops a space
\thanks{$^{1}$CNNP Lab (www.cnnp-lab.com), School of Computing, Newcastle University, Newcastle upon Tyne, NE4 5TG, United Kingdom}%
\thanks{$^{2}$Faculty of Computing and Engineering,  Atlantic Technological University, Letterkenny, F92 FC93, Ireland}%
\thanks{$^{3}$Faculty of Medical Sciences, Newcastle University, Newcastle upon Tyne, NE2 4HH, United Kingdom}%
\thanks{$^{4}$UCL Queen Square Institute of Neurology, Queen Square, London, WC1N 3BG, United Kingdom}%
\thanks{Please address correspondence to: {\tt\small emmanuel.molefi@ncl.ac.uk} or {\tt\small yujiang.wang@ncl.ac.uk}}
}
\begin{document}

\bstctlcite{IEEEBSTCTL:BSTcontrol}

%-------------------------------------------------------------------------%

\maketitle
\thispagestyle{empty}
\pagestyle{empty}

%-------------------------------------------------------------------------%
\begin{abstract}
Ultradian rhythms -- quasi-rhythmic fluctuations in behavior and physiology with periods shorter than 24 hours -- are observed across various organisms, including humans. Despite their role in key biological processes such as sleep architecture and hormone regulation, their underlying mechanisms remain poorly understood. Here, we leveraged wearable sensor technology for continuous monitoring of physiological signals in 16 healthy participants over two weeks. By systematically removing circadian and longer-scale rhythms, we isolated ultradian dynamics and modeled them using the Hankel Alternative View of Koopman (HAVOK) framework, a data-driven approach based on Takens’ embedding theorem and Koopman operator theory. This allowed us to characterize ultradian rhythms as an intermittently forced linear system and distinguish between regular oscillatory behavior and more complex dynamics. Across participants, ultradian fluctuations were well-described by the HAVOK model, with intermittent forcing consistently observed. The model demonstrated strong forecasting accuracy, with root mean squared error (RMSE) of 0.0315 ± 0.02, 0.0306 ± 0.02, and 0.0218 ± 0.02 in the leading time-delay coordinates. Notably, a significant sex difference in model rank (\textit{z} = -2.06, \textit{p} = 0.0396) suggests that sex hormones may play a key role in ultradian dynamics. These findings provide evidence for intermittently forced linear systems as a useful framework for understanding ultradian rhythms and their regulation. 
\newline

\indent \textit{Clinical relevance}— Disruptions in ultradian rhythms are linked to neurological and psychiatric disorders. Identifying their key driver dynamics could inform chronotherapy and biomedical interventions, offering new strategies for regulation in health and disease.
\end{abstract}
%-------------------------------------------------------------------------%

%-------------------------------------------------------------------------%
\section{INTRODUCTION}

Biological rhythms are everywhere in nature, from the ebb and flow of ocean tides affecting marine organisms to the yearly hibernation cycles of some animals. These rhythms help organisms to structure their activities in time and stay in sync with their environment. 

Humans also rely on various biological rhythms to regulate their bodies and behaviors. A well-studied example is the 24~hour sleep-wake cycle, but we also follow rhythms on other timescales. For example, our hormone levels, the so-called rest-activity cycle, and our sleep architecture follow a repeating cycle of a few hours. These shorter cycles, known as ultradian rhythms (lasting less than 24 hours), are often less well-studied, but altered or disrupted ultradian rhythmicity is a key feature of various disease states (e.g., depression~\cite{hall1996ultradian}, schizophrenia~\cite{scott2023twelve}, and epilepsy~\cite{thornton2024diminished}). 

Despite their importance, the mechanisms and dynamics underlying ultradian rhythms remain poorly understood. For example, the ultradian rhythm of cortisol -- a key stress hormone -- is believed to result from concentrated pulses of hormone secretion every 100 minutes or so~\cite{benz2019duration}. In contrast, the transitions between sleep stages are often described as critical transitions between distinct ``states''. Here, we therefore seek to analyze ultradian rhythms in a dynamical framework. 

Ultradian rhythms can be measured in a variety of ways, depending on the physiological variable of interest. In this study, we opted for wearable technology to allow for continuous, long-term monitoring of multiple physiological signals in real-world conditions. We obtained data from healthy individuals over approximately two weeks to examine ultradian patterns. By systematically removing circadian influences and longer timescales, we isolated and analyzed the ultradian components of these physiological signals. 

Our objective is to represent ultradian dynamics as an intermittently forced linear system, applying the HAVOK~\cite{brunton2017chaos} framework -- a data-driven method for dynamical systems modeling that leverages early concepts from Takens' embedding theorem~\cite{takens1981detecting} and Koopman operator theory~\cite{koopman1931hamiltonian}. Takens' embedding theorem has been widely used to analyze and characterize dynamical systems. Takens~\cite{takens1981detecting} showed that time-delay embedding provides a reconstructed attractor that is diffeomorphic to the original attractor of the dynamical system. Modern research has revealed a connection between this powerful technique and the Koopman operator -- which provides a linear representation of nonlinear dynamics~\cite{mezic2004comparison}. The HAVOK framework relies on this connection, showing that dynamics of a nonlinear system can be captured by a linear model with intermittent forcing. 

Accordingly, this approach allows us to distinguish the types of linear dynamics that would be generated by a simple/regular oscillator \textit{vs.} more complex dynamics, which are modeled as a forcing term. For the ultradian fluctuations we observe, we are essentially asking the question of what types of simple oscillator would generate them -- based on analysis of their modeled dynamical behavior -- and what occasional forcing do these oscillators experience to produce the observed quasi-periodic fluctuations. 

Understanding the linear dynamics and intermittent forcing of ultradian rhythms could have significant implications. By identifying the key oscillators and drivers of ultradian fluctuations, we may open the door to targeted interventions that help regulate these rhythms, offering potential benefits for conditions where ultradian disruption plays a role, such as mood disorders, epilepsy, and other neurological conditions. 
%-------------------------------------------------------------------------%

%-------------------------------------------------------------------------%
\section{METHODS}

%-------------------------------------------------------------------------%
\subsection{Participants and Data Acquisition}

This study was approved by the Newcastle University Ethics Committee (23-032-WAN), and conformed to the Declaration of Helsinki standards for human research. Participants were 16 healthy volunteers (mean age $\pm$ S.D. = 34.7 $\pm$ 10.0 years, age range = 21-66 years, 11 females) recruited from the staff and student population at Newcastle University. All participants provided written informed consent at the start of their in-person wearable devices setup visit. We measured heart rate (HR), peripheral skin temperature (Temp), and activity counts (Act) using the Empatica EmbracePlus wristband~\cite{gerboni2023prospective}, and performed continuous glucose monitoring (CGM) with the FreeStyle Libre 2 system~\cite{fokkert2017performance} over the course of approximately 2 weeks (mean 12.9 days, range 7-16 days). 
%-------------------------------------------------------------------------%

%-------------------------------------------------------------------------%
\subsection{Data Preprocessing and Ultradian Rhythm Extraction}

The obtained Empatica EmbracePlus and FreeStyle Libre 2 data were preprocessed and analyzed using scripts written in MATLAB R2024b (The MathWorks, Inc., Natick, MA, USA). First, raw signals (i.e., HR, Temp, Act, and CGM) were resampled to 15-min time intervals -- calculating a mean for each interval. Any missing data of less than 24~hrs were filled using linear interpolation. Any longer periods with missing data were discarded, only retaining the longest continuous segment of data. 

\begin{figure*}[ht!]
  \centering
  \includegraphics[width=\linewidth]{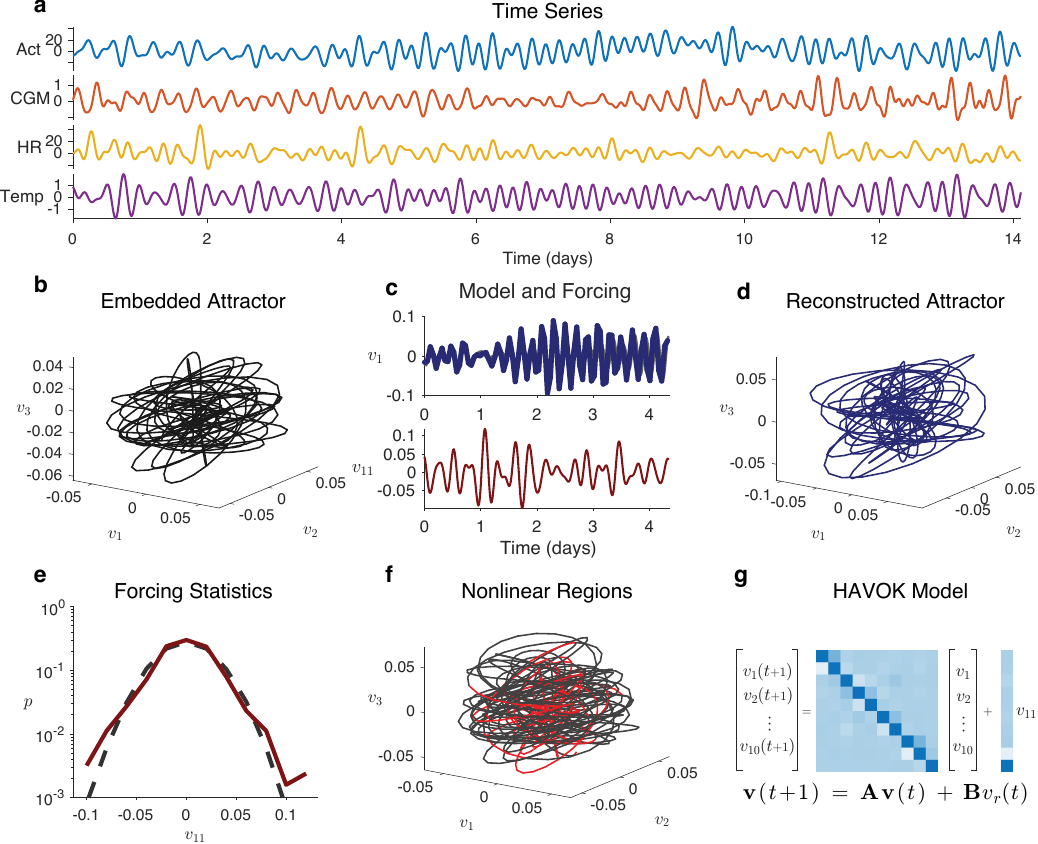}
  \caption{Multimodal modeling of ultradian rhythms as an intermittently forced linear system. (\textbf{a}) Observed physiological time series measurements from one example participant; we consider these multivariate data in the form of a matrix ${\bf{X}}$, and construct a Hankel matrix ${\bf{H}}$ -- which in turn is used to compute the SVD to obtain hierarchically ordered eigen time series that produce a delay-embedded attractor (\textbf{b}). For example data in (a), a HAVOK linear model is obtained on the first $r - 1$ delay coordinates forced by the last delay coordinate $v_r$. Using a HAVOK model with forcing from $v_{11}$, we perform reconstruction of $v_1$ (\textbf{c}) and embedded attractor in $(v1, v2, v3)$ (\textbf{d}). In (\textbf{e}), a PDF of the forcing term in $v_{11}$ is compared with that of a normal distribution. (\textbf{f}) The delay embedded attractor colored by the activity of the forcing term $v_{11}$ reveals trajectories corresponding to regions when the forcing is inactive (gray), indicating a well-approximation of the dynamics by Koopman linear dynamics; and when active (red), denoting intermittent forcing. (\textbf{g}) HAVOK model obtained on the time-delay coordinates of ultradian rhythms; the model takes the form of a difference equation given by~(\protect\ref{eq:HAVOK}).}
  \label{fig:UltradianRhytmicityModeling}
\end{figure*}

For ultradian rhythm extraction, time series decomposition was performed using singular spectrum analysis (SSA)~\cite{nina2013singular} with a $lag = 288$ to identify periodic components. Note, we centered our data by subtracting column means before performing SSA. To classify ultradian components, we computed the multitaper power spectral density (PSD) estimate~\cite{thomson1982spectrum} of each component to estimate the period. We then summed components with a peak period between 2 and 12~hrs inclusive to obtain the ultradian rhythm. Following extraction of ultradian rhythms for each time series signal (i.e., HR, Temp, Act, and CGM), we constructed an $n \times m$ data matrix ${\bf{X}}$, the columns ($m$) of which represented the time points; and rows ($n$) denoted the different data streams~(Fig.~\ref{fig:UltradianRhytmicityModeling}a). 
%-------------------------------------------------------------------------%

%-------------------------------------------------------------------------%
\subsection{Hankel Alternative View of Koopman (HAVOK) Analysis}

To decompose ultradian rhythms into a linear system with intermittent forcing, we performed HAVOK analysis~\cite{brunton2017chaos}. In brief, the HAVOK method builds models via time-delay embeddings; considering a univariate time series $x(t)$, eigen-time-delay coordinates are obtained by computing the singular value decomposition (SVD) of the Hankel matrix ${\bf{H}}$: 

\begin{equation} 
\label{eq:Hankel}
{\bf{H}} = \left[ {\begin{array}{*{20}{c}} {x({t_1})} & {x({t_2})} & \cdots & {x({t_p})} \\ \\ {x({t_2})} & {x({t_3})} & \cdots & {x({t_{p + 1}})} \\ \\ \vdots & \vdots & \ddots & \vdots \\ \\ {x({t_q})} & {x({t_{q + 1}})} & \cdots & {x({t_m})} \\ \end{array}} \right] = {\bf{U}}\Sigma {{\bf{V}}^{*}},
\end{equation}

where ${\bf{U}}$ and ${\bf{V}}$ are matrices whose columns are hierarchically ordered by how much variance they capture in the columns and rows of ${\bf{H}}$, respectively. Note that the Hankel matrix in~(\ref{eq:Hankel}) is constructed from time-shifted copies of the measurement $x(t)$. Performing the SVD as shown yields an optimal low-rank approximation to ${\bf{H}}$ -- keeping the leading $r$ columns of ${\bf{U}}$ and ${\bf{V}}$; more importantly, this obtained low-rank approximation provides a data-driven measurement system that is approximately invariant to the Koopman operator. Hence,~(\ref{eq:Hankel}) may be expressed with the Koopman operator ${\cal K}$, giving:  

\begin{equation}
{\bf{H}} = \left[ {\begin{array}{*{20}{c}} {x({t_1})} & {{\cal K}x({t_1})} & \cdots & {{{\cal K}^{p - 1}}x({t_1})} \\ \\ {{\cal K}x({t_1})} & {{{\cal K}^2}x({t_1})} & \cdots & {{{\cal K}^p}x({t_1})} \\ \\ \vdots & \vdots & \ddots & \vdots \\ \\ {{{\cal K}^{q - 1}}x({t_1})} & {{{\cal K}^q}x({t_1})} & \cdots & {{{\cal K}^{m - 1}}x({t_1})} \\ \end{array}} \right].
\end{equation}

Taking this perspective, including early evidence providing a link between the Koopman operator and Takens' embedding~\cite{mezic2004comparison}, the HAVOK architecture produces a linear HAVOK model on the variables in ${\bf{V}}$~\cite{brunton2017chaos}.

As our data was multivariate in nature, we constructed our Hankel matrix ${\bf{H}}$ with multivariate time-delay embedding as described in~\cite{dylewsky2022principal}, using~(\ref{eq:Hankel2}) in lieu of the approach taken in~(\ref{eq:Hankel}). Here $x(t) \in {\rm I\!R}^n$ with the data matrix being 

\begin{equation}
\mathbf{X} = \begin{bmatrix}
| & | &  & |\\
{x}(t_1) & {x}(t_2) & \cdots & {x}(t_m)\\
| & | &  & |
\end{bmatrix}.
\end{equation}

\begin{equation}\label{eq:Hankel2}
\mathbf{H} = \begin{bmatrix}
| & | &  & |\\
{x}(t_1) & {x}(t_2) & \cdots & {x}(t_p)\\
| & | &  & |\\
| & | &  & |\\
{x}(t_2) & {x}(t_3) & \cdots & {x}(t_{p + 1})\\
| & | &  & |\\
\vdots         &  \vdots        &  \ddots&  \vdots\\
| & | &  & |\\
{x}(t_q)& {x}(t_{q + 1})  &  \vdots	& {x}(t_m) 	\\
| & | &  & |\\
\end{bmatrix}
= {\bf{U}}\Sigma {{\bf{V}}^{*}}.
\end{equation}

Following computation of the SVD of ${\bf{H}}$ to obtain eigen-time-delay coordinates, we constructed a linear regression model on the first $r - 1$ variables in ${\bf{V}}$ with linear forcing provided by the last variable $v_r$: 

\begin{equation}\label{eq:HAVOK}
{\bf{v}}(t + 1) = {\bf{Av}}(t) + {\bf{B}}{v_r}(t),
\end{equation}

where ${\bf{v}}{(t)} = {[ {\begin{array}{*{20}{c}} {{v_1}} & {{v_2}} & \cdots & {{v_{r - 1}}} \end{array}} ]^T}$ is the leading $r - 1$ eigen-time-delay coordinates. 
%-------------------------------------------------------------------------%

%-------------------------------------------------------------------------%
\subsection{Statistical Analysis}

All statistical analysis were performed in MATLAB R2024b. Probability density functions (PDFs) were computed for performing comparisons between the forcing term $v_r$ and a normal distribution. To assess model generalization to validation data, we computed the  root mean squared error (RMSE). Sex differences in model hyperparameters were compared using a two-sided Wilcoxon rank sum test at the $p < 0.05$ significance level. All data in the graphs represent median and quartiles. 

%-------------------------------------------------------------------------%

%-------------------------------------------------------------------------%
\section{RESULTS}

%-------------------------------------------------------------------------%
\subsection{Multimodal decomposition of ultradian dynamics into a linear dynamical system with forcing}

We observed that ultradian rhythms derived from multimodal data may be represented as an intermittently forced linear system. Fig.~\ref{fig:UltradianRhytmicityModeling} illustrates the HAVOK analysis for one example participant. We show a delay-embedded attractor~(Fig.~\ref{fig:UltradianRhytmicityModeling}b) obtained from eigen time series (given by the columns of ${\bf{V}}$) that characterizes the measured ultradian dynamics. In Fig.~\ref{fig:UltradianRhytmicityModeling}c, we demonstrate a reconstruction of the dominant time-delay coordinate $v_1$ using a linear HAVOK model with forcing from $v_{11}$. We find that the attractor dynamics of the observed ultradian patterns (shown in gray; top panel) are well-captured by a HAVOK model (shown in blue); moreover, the HAVOK model can be used to reconstruct the embedded attractor, as shown in Fig.~\ref{fig:UltradianRhytmicityModeling}d. This suggests that ultradian dynamical patterns are well-approximated by Koopman linear dynamics with intermittent forcing.

Fig.~\ref{fig:UltradianRhytmicityModeling}e compares the PDFs of the forcing term $v_{11}$ (dark red line) and the normal distribution (dashed gray line), showing that the forcing statistics are non-Gaussian, denoting rare forcing events in ultradian variation; importantly, this forcing signal suggests that there may be a simple oscillator that governs ultradian rhythms.

The colors of the time-delay embedded attractor presented in Fig.~\ref{fig:UltradianRhytmicityModeling}f reveal to us the activity of the forcing term between well approximated Koopman linear dynamics (gray) and where ultradian variation is nonlinear (red). Fig.~\ref{fig:UltradianRhytmicityModeling}g displays the structure of the HAVOK model for the observed ultradian rhythms; this model portrays a dominant off-diagonal structure in agreement with the findings in~\cite{brunton2017chaos}. 
%-------------------------------------------------------------------------%

%-------------------------------------------------------------------------%
\subsection{Characterizing ultradian rhythms with forcing}

\begin{figure}[th]
  \centering
  \includegraphics[width=\linewidth]{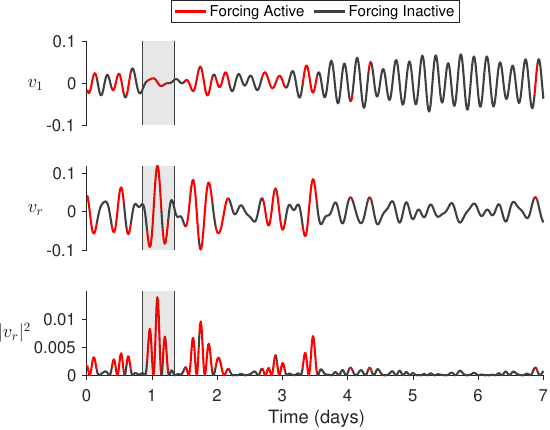}
  \caption{The $v_1$ and $v_r$ eigen time series for ultradian rhythms; the forcing signal $v_r$ is also shown as $|v_r|^2$ for pronounced visualization. The forcing is active (red line) when the threshold $|v_r| > 0.0011$, and inactive (gray line) below this threshold; when the forcing is inactive, the dynamics are well-approximated by the linear Koopman model, and when active, suggests a rare intermittent event. The annotated region (gray background) highlights a particularly critical event with large forcing, denoting a rarer event.}
  \label{fig:ForcingSignature}
\end{figure}

Fig.~\ref{fig:ForcingSignature} shows the $v_1$ eigen time series for ultradian rhythmicity zoomed in to the first 7 days. This trajectory has been color-coded according to the activity of the external forcing with respect to a threshold value (in this case, $r = 11$ and the threshold is $0.0011$): when the forcing is active (red line), it suggests the presence of nonlinear dynamics that denote a rare forcing event; and when inactive (gray line), the rhythms are well-described by Koopman linear dynamics. 
%-------------------------------------------------------------------------%

%-------------------------------------------------------------------------%
\subsection{Prediction of ultradian dynamics}

\begin{figure}[th]
  \centering
  \includegraphics[width=\linewidth]{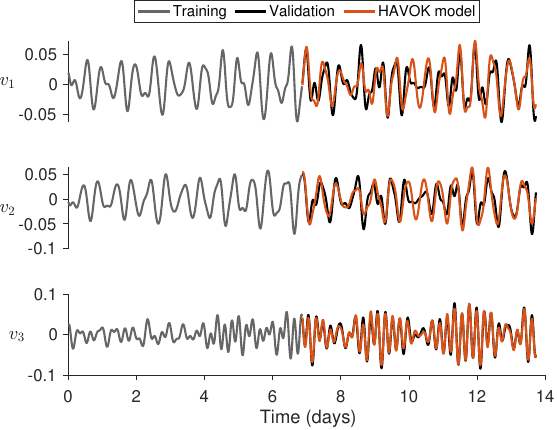}
  \caption{HAVOK predictions of the first three time-delay-coordinates $(v1, v2, v3)$ for ultradian dynamics obtained from multimodal data from one example participant. The HAVOK model obtained from training data (initial 7 days; gray) was evaluated on a new validation trajectory (black) in the next 7 days.}
  \label{fig:ModelValidation}
\end{figure}

\begin{figure}[th]
  \centering
  \includegraphics[width=\linewidth]{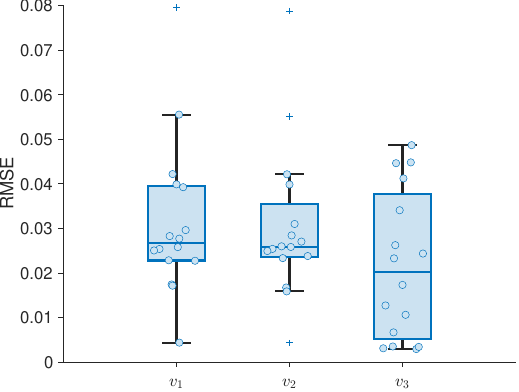}
  \caption{RMSE results summarizing HAVOK prediction errors of the first three time-delay-coordinates $(v1, v2, v3)$ against validation data from future unseen trajectories of ultradian dynamics for all participants.}
  \label{fig:RMSE}
\end{figure}

Extending our analysis to prediction of future ultradian dynamics, we found that a HAVOK model trained using data from the initial 7 days, made predictions that closely match the data in the following 7 days (Fig.~\ref{fig:ModelValidation}). In this example, the HAVOK model generalizes particularly well to eigen-time-delay coordinate $v_3$ of the unseen trajectory. Fig.~\ref{fig:RMSE} demonstrates that eigen-time-delay coordinate $v_3$ was indeed generally modeled well across all participants. The average RMSE on the validation (test) data was $0.0315 \pm 0.02$, $0.0306 \pm 0.02$, and $0.0218 \pm 0.02$ (mean $\pm$ S.D.) in $(v_1, v_2, v_3)$, respectively -- across all participants. Overall, the HAVOK models demonstrated low RMSE values ($< 0.08$), indicating good generalization to unseen trajectories. 
%-------------------------------------------------------------------------%

%-------------------------------------------------------------------------%
\subsection{Sex differences in model hyperparameters}

\begin{figure}[th]
  \centering
  \includegraphics[width=\linewidth]{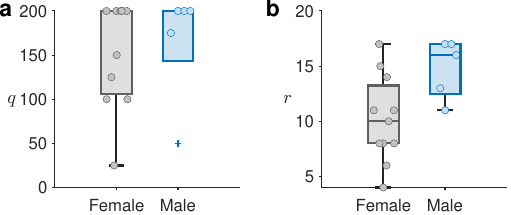}
  \caption{Comparison of HAVOK models hyperparameters number of delays $q$ (\textbf{a}) and model rank $r$ (\textbf{b}) between female and male participants.}
  \label{fig:Sex_and_Hyperparameters}
\end{figure}

We compared the HAVOK model hyperparameters $q$ (number of delays) and $r$ (model rank) between female and male participants~(Fig.~\ref{fig:Sex_and_Hyperparameters}). Although there was no difference in hyperparameter $q$ between the sexes ($z = -0.25, \ p = 0.8027$), there were pronounced sex differences in hyperparameter $r$. HAVOK models from male participants were characterized by a higher rank~($z = -2.06, \ p = 0.0396$; Fig.~\ref{fig:Sex_and_Hyperparameters}b). 

As an aside, we found that both sexes generally had $q$ of at least $100$~(Fig.~\ref{fig:Sex_and_Hyperparameters}a); this echoes the notion that higher values of $q$ produce good HAVOK models~\cite{brunton2017chaos,hirsh2021structured}. 
%-------------------------------------------------------------------------%

%-------------------------------------------------------------------------%
\section{DISCUSSION}

Our study provides initial evidence to suggest that HAVOK models using multimodal data can capture ultradian dynamics; and thus represent an important tool to study ultradian rhythmicity. In particular, we have demonstrated that with an approximately Koopman-invariant measurement system, ultradian rhythms can be decomposed into linear dynamics with intermittent forcing. Moreover, we showed that these Koopman linear dynamics have the capability to generate future ultradian variation trajectories, showing their ability to generalize. 

We consistently observed HAVOK models that revealed a good description of the data as depicted in Fig.~\ref{fig:UltradianRhytmicityModeling}c (top panel); furthermore, the obtained forcing statistics of $v_r(t)$ portrayed a non-Gaussian distribution (Fig.~\ref{fig:UltradianRhytmicityModeling}e), with long tails indicating the presence of rare intermittent events. It is possible that this rare-event forcing implies activation -- and thus existence -- of ultradian stimuli. These could reflect endogenous processes (e.g., hormone secretion) or exogenous stimuli, such as acute stressors.  In light of this, it is notable that recent data~\cite{zhu2024evidence} provide evidence for ultradian rhythms at the molecular level in humans on which to integrate our findings. 

Examining the annotated forcing activity in tandem with the time history in the dominant eigen-time-delay coordinate $v_1$ (Fig.~\ref{fig:ForcingSignature}), it appears that perturbed ultradian rhythms correspond to large forcing. When these rhythms are in the expected range of variation, the forcing signature is small. Interestingly, this forcing signature is larger in the annotated region~(Fig.~\ref{fig:ForcingSignature}) compared to other forcing activations, suggesting that the forcing activity could reveal to us not only of a rare event occurring but also indicate differential intensity of the event -- that is, signify the importance of altered dynamics. Importantly, this could provide warning signs about deteriorating ultradian rhythmicity. 

A notable finding from our study is the demonstration that HAVOK models are potentially predictive of ultradian dynamics~(Figs.~\ref{fig:ModelValidation} and~\ref{fig:RMSE}). This indicates that it may be possible to rely on ultradian dynamical patterns from previous measurements to inform the future. We~\cite{thornton2024diminished} and others~\cite{armitage1999biological} have previously found disrupted ultradian rhythms in epilepsy and depression, respectively, using neural recordings. Here, by revealing the potential for HAVOK models to forecast ultradian variation, our findings provide novel insight to guide future development of targeted therapeutic interventions -- such as chronotherapy and neuromodulation -- for improved health outcomes in these disabling conditions. 

When comparing model hyperparameters between female and male participants, we observed that both sexes generally had higher values for hyperparameter $q$ (number of delays)~(Fig.~\ref{fig:Sex_and_Hyperparameters}a). That there was no observed difference in $q$ between the sexes is to be expected, given that model performance has been found to improve with higher number of delays~\cite{brunton2017chaos,hirsh2021structured}. Interestingly, we observed that models from female participants performed better in modeling ultradian dynamics with higher rank $r$ compared to male participants~(Fig.~\ref{fig:Sex_and_Hyperparameters}b), suggesting that the expression of ultradian rhythms may be affected by differences in sex hormones; and that understanding these differences could provide valuable insights into ultradian biology. 

Overall, this study has provided insights that could lead to therapeutic strategies for, e.g., epilepsy, depression, and schizophrenia. For instance, in epilepsy, our multimodal HAVOK modeling framework could be used to characterize interactions between seizure occurrence and disrupted ultradian variation, whereby the HAVOK forcing signature would signal an impending seizure, and in turn inform time-based interventions that target and promote healthy ultradian dynamics in affected individuals. 

In the field of psychiatry, we envisage that this framework may provide a forcing signal that is potentially predictive of a mental health episode based on long-term trends of ultradian dynamical patterns; thus, facilitate chronotherapeutic interventions. Specifically, the forcing signature could provide early indications of abnormally elevated mood or extreme mood changes in bipolar disorder, and unveil the impact of a delayed cortisol rhythm observed in adults with attention-deficit hyperactivity disorder (ADHD)~\cite{baird2012adult}, for example. Moreover, the forcing activity may provide insights into how long it takes for altered ultradian dynamics to recover and return to baseline variation following manic and depressive bipolar episodes, or whether the dynamics exhibit persistent abnormality. Furthermore, the rare-event forcing observed could offer insights into how interventions for mental health conditions work, in particular to normalize ultradian variation. Importantly, this highlights how our analytical approach may be relevant for a broader spectrum of neurological and psychiatric disorders. 

Finally, future work should investigate if circadian rhythmicity could also be characterized by linear HAVOK models; and importantly, if ultradian forcing agrees with actually observed ultradian dynamics. In other words, we could establish a hierarchy of rhythms and understand how different timescales interact; ultimately for therapeutic benefit.
%-------------------------------------------------------------------------%

%-------------------------------------------------------------------------%
\section{CONCLUSIONS}

The findings from our study demonstrate that ultradian rhythms can be characterized as an intermittently forced linear system using the HAVOK framework -- a data-driven procedure based on Takens' embedding and Koopman theory. We note that this multimodal modeling framework could be used for other biological rhythms (e.g., circadian, infradian, etc.), including other modalities (e.g., neural recordings). 
%-------------------------------------------------------------------------%

\addtolength{\textheight}{-12cm}

%-------------------------------------------------------------------------%
\section*{ACKNOWLEDGMENT}
Data collection for the study was in part supported by the EPSRC CloseNIT network (EP/W035081/1). P.N.T. and Y.W. are both supported by UK Research \& Innovation (UKRI) Future Leaders Fellowships (MR/T04294X/1, MR/V026569/1).
%-------------------------------------------------------------------------%

%-------------------------------------------------------------------------%
\bibliographystyle{IEEEtran}
\bibliography{references}
%-------------------------------------------------------------------------%

\end{document}